\documentclass{aa}
\usepackage{amssymb,graphicx}

\newcommand{\gaeq}{\hbox{\raise.5ex\hbox{$>$}
    \kern-1.1em\lower.5ex\hbox{$\sim$}}}

\newcommand{\laeq}{\hbox{\raise.5ex\hbox{$<$}
    \kern-1.1em\lower.5ex\hbox{$\sim$}}}

\begin{document}

   \thesaurus{11     
              ( 
               )} 

   \title{The Surroundings of Disturbed, Active Galaxies}

   \author{Halton Arp}

   \institute{Max-Planck-Institut f\"ur Astrophysik, 85740 Garching, Germany}
     
    \offprints{H. Arp, arp@mpa-garching.mpg.de}
    
     \date{Received April 1999}

        \maketitle

        \begin{abstract}

The brightest apparent magnitude examples of ultra luminous infrared galaxies 
(ULIRG's) are studied here in X-ray, optical, infrared and radio wavelengths. 
It is is found that they have associated material reaching out to apparent 
diameters of the order of a degree on the sky. 
 
Gas, dust, X-ray material and quasars appear to be
ejected from the active nuclei with all objects nearer than 
their redshift distances.


\end{abstract}

\section{Introduction}
    
The neighborhoods of the most active ULIRG's, Mark273, Mark231, Arp 220 and 
NGC6240 plus the radio galaxy 3C31$/$NGC383 are examined for associated 
objects.

\section{Markarian 273}

     A strong jet emerges S from the disturbed central regions of this ULIRG. Fig.1
shows that both Mark 273 (z = .038) and the compact object Mark 273x (z =
.458) are strong
X-ray sources. The deep optical image in Fig. 2 illustrates the conclusion by Xia et
al. (1999) that "Mark 273x is at the tip of the plume" reaching NE from Mark 273.

     Fig.3 addresses the question of whether there is any X-ray connection
between Mark 273 and the AGN to the NE. The outer isophotes do not appear as
rounded as one would expect from accidental adjacency of unrelated,
symmetrical sources. (See for example the contouring of two adjacent,
unrelated images in Arp 1998a, Fig. 1-12). In addition the bright central regions of 
Mark 273 are conspicuosly extended toward Mark 273x.

     In fact what we see is a compact optical object of R = 19.6 mag. and redshift z
= .458 connected back to a disturbed  Seyfert 2 of V = 14.9 mag. by both optical
and X-ray emitting material. Mark 273x is the same kind of quasar-like object
which was physically associated with somewhat brighter Seyfert galaxies at a 7.5
sigma level (Radecke 1997; Arp 1997). Also there is a recent precedent for a
quasar of z = 2.15 at the tip of an X-ray/optical filament emerging along the
minor axis of the active galaxy NGC3628 (Flesch and Arp 1999). 

     The luminosity of the quasar-like Mark 273x as derived by Xia et al.
(1998) when they believed it to be connected to Mark 273 was 
M$_R$ =-17.6 mag. But we will argue later that Mark 273 
itself is quasar-like in that it has an intrinsic redshift and should be moved of 
the order of 5 magnitudes (in modulus) closer.

     Smoothing and contouring the lower resolution, PSPC X-ray measures
produces Fig. 4. It is seen that the outer contours of Mark 273x verify the
elongation to the NE, along a line back toward the central galaxy. Regardless of
whether this x-ray connection is due to a jet or to some merger related activity it
does establish the fact that the two objects are at the same distance. There is,
however a clear X-ray jet visible in the PSPC measures. It comes out of Mark 273
to the SE and is delineated even in the 3,4 and 5 sigma contours (also more
weakly in the opposite direction).

     There is no reason not to have jets coming out of active galaxies in different
directions. In fact there is considerable evidence for lines of X-ray sources 
emerging from Seyfert and other active galaxies (Arp 1996;1997). Why those
apparent ejections take place in roughly orthogonal directions, as here in Mark
273, there seems to be no ready explanation, but it does seem to be an empirical
result. As we shall see there is evidence for X-ray material and objects coming
out further along these same principal directions.

     As for the relation of Mark 273x to Mark 273, the ejection of higher redshift
quasars and quasar-like objects has been argued for more than 30 years
(see Arp 1983; 1998a).
But the interesting additional information given by these {\it disturbed}, active
galaxies is that when quasars do not come out along relatively unimpeded along
the minor axis, that they then interact with the material of the ejecting galaxy and
disrupt it, giving rise to disturbed morphologies, entrainments and 
fragmentation (Arp 1999a).

\section{Markarian 273 Radio Features}
     It is most interesting to consider the information in other wavelengths, not
only to confirm the features we have already seen but also to try to understand
the physcial processes involved in the ejections, if that is the mechanism for
producing these associations. 

\vskip .5cm

     In Fig.5 we see the map of the Mark 273 region in continuum radio
wavelengths. The most prominent feature is a broad ejection coming out NW and
SE, agreeing closely with the direction of the X-ray jet. (Close to the nucleus,
agreement between X-ray and radio jets is characteristic of very active radio
galaxies like Virgo A and Centaurus A). In the counter jet direction the radio
extension is longer but weaker. Overall, there seems to be confirmation that both
X-ray and radio emitting material are coming out of Mark 273 at position angles
of about 130 and 310 degrees.

     The next most conspicuous feature in Fig. 5 is the very strong radio source
about 5.5 arcmin SW of Mark 273. This radio source is a close double (about 1.1
arcsec separation and p.a. about 20 deg when viewed on the VLA, high
resolution mode, FIRST). Doubleness of companion radio and X-ray sources will
be noted as we explore other active galaxy neighborhoods. But of considerable
significance for the ejection hypothesis for Mark 273x is the fact that it is itself 
a strong radio source, and that it is {\it aligned, as exactly as can be measured},
across Mark 273 with the strong radio source to the SW. 

     Fainter radio extensions than contoured in the preceding Figure are shown
here in Fig. 6. They extend both NW and SE generally along the direction of the 
radio and X-ray jets. (The arrow in Fig. 6 marks the X-ray jet pictured in Fig. 4). 
Surprisingly there are two low surface brightness radio filaments extending from 
Mark 273 and ending on two very strong X-ray sources (in Fig. 7 they are marked 
as Nos. 21N and 38 and are at 6.2 and 6.6 arcmin respectively from Mark 273). 
They are marked with x's in Fig. 6. One is a quasar of z = .941 and the other is a 
blue stellar object (BSO) with a similar apparent magnitude, of 18.1. The latter 
will almost certainly turn out to be a second quasar in a pair. In the deep red 
exposure of Fig. 2 there are twin, low surface brightness {\it optical} filaments 
coming out of the S end of Mark 273. (Seen better in deeper image in J.Hibbard's
home page.) They appear to lead into the radio filaments which are observed in 
Fig. 6. In summary, it appears that in all the wavelengths that can be checked, 
radio, optical and X-ray, there is a connection of these two strong X-ray sources 
back to the center of Mark 273. 

     Because of the apparent quasar nature and consequent high redshift of these 
two strong X-ray sources it is customary to ask whether they could be 
accidentally projected background sources. We can answer this suggestion by
noting that the strength of these X-ray sources: Mark 273x = 8.3 cts/ks, z = .941
quasar = 37.7 cts/ks and 18.1 BSO = 21.2 cts/ks indicates that they would have a
chance of being background sources falling as close as they do to Mark 273 of p
= .003, .01 and .04. Altogether getting just these three objects this close to an
arbitrary active galaxy would only be about one chance in a million. Xia et al. 
(1999) note in passing that it is interesting that "... the X-ray companions of the 
three nearest ULIG's (Arp 220, Mrk 273 and Mrk 231) are all background 
sources...".

 \section{X-ray field around Mark 273}

     Since X-ray sources and quasars have been  associated with Seyfert galaxies
out to distances of nearly a degree it is of interest to see what the full PSPC 
field of ROSAT around mark 273 looks like.  

\vskip 1.0cm

     The strong sources that we have been discussing  are seen close to the
center of Fig. 7. The two probable quasars are seen to be exceptionally strong
X-ray sources at 21 and 38 cts/ks, much stronger even than Mark273 and 273x
which have 13 and 8 cts/ks. As we go off-axis the point spread of the images
enlarges so only the brighter sources are registered. It is clear, however, that
sources of all brightness are distributed in an elongated pattern, roughly NE to
SW through Mark 273. This line is close to the direction of Mark 273x and the
strong, double radio source pictured in Fig. 5 (p.a. $\sim$ 40 deg.). 

     Of course the radio and X-ray ejections which lead toward the two strong
X-ray sources (21N and 38) are in an almost orthogonal direction to this (p.a
$\sim$ 140 deg.). They are suggested to be connected with the the strong X-ray
sources 16 and 30 which lie further out in this direction. As Table 1 shows, both
the 16 and 30 cts/ks X-ray sources are optically identified as blue stellar objects
(BSO's) and therefore are most probably a quasar pair like so many of the
X-ray sources paired across Seyfert galaxies. 
     
     Ejection in more than one direction from active galaxies is commonly
observed (see Arp 1996; 1997a; Komossa and Schulz 1998). {\it The most 
important aspect of Fig. 7, however, is that numerous X-ray sources in this field 
are aligned preferentially across Mark 273, confirming that they are predominantly 
not background.}

\vskip 2ex

\centerline{TABLE 1}       
             
\centerline{Bright X-ray sources in Mark 273 field}

\vbox{\tabskip=0pt \offinterlineskip
\def\tr{\noalign{\hrule}}
\halign to \hsize{\tabskip=1em minus 1em&
\strut#\hfill&#\hfill&#\hfill&#\hfill&#\hfill&#\hfill&#\hfill
\tabskip=0pt\cr
 Cts &R.A.~~(2\rlap{000)}&\hfill Dec.&~~E&O-E&ID&Remarks\cr
\tr
  49 &13 41 36.2&~ 55 14 37 & 16.93 & 1.99 & c.g. & \cr
  26 &13 42 06.3&~ 56 39 15 & 16.52 & 1.20 & BSO & 47"off pos \cr
  16 &13 42 21.6&~ 56 14 50 & 19.88 &  ~.73  & BSO & \cr
  18 &13 44 13.5&~ 56 30 34 & 18.30 &  ~.34  & BSO & \cr
  21S&13 44 57.3&~~55 28 22 & 18.9   &  ~.85  &  ?     & extend?\cr
  38  &13 44 47.3&~~55 46 56 &  17.76 &  ~.77  & QSO & z = .941\cr 
  13  &13 44 42.1&~~55 53 13 &  14.9*  &  ~.77* & S2    & Mark 273\cr 
   8   &13 44 47.1&~~55 54 10 &  19.25 & 1.78  & c.g.  & Mark 273x\cr
  21N&13 45 12.0&~~55 47 59 &  18.1   &  ~.57   & BSO \cr  
  29  &13 45 33.4&~~55 24 06 &  19.26  &  ~.66   & BSO\cr
   8   &13 46 06.7&~~56 04 29 &  19.08  &  ~.28   & BSO \cr
  30  &13 46 38.9&~~55 27 09 &  19.44  &  ~.32   & BSO \cr
  48  &13 46 59.1&~~56 07 04 &   18.21 & 1.49  & BSO \cr
}} 

* B and B-V magnitudes

\section{Infrared sources aligned with Mark 273}

In order to investigate the Mark 273 surroundings in longer wavelengths the
Simbad on-line catalog was consulted. It was noticed immediately that there was
a conspicuous string of what turned out to be infrared sources coming
generally SW from the ULIRG galaxy. Fig. 8 shows a plot of nearly the same
sized area as pictured in the preceding PSPC X-ray map. {\it Only infrared point
sources with 100 micron fluxes greater than .5 Jy have been plotted.} 
The sources are as listed in the IRAS faint source catalog of 173,044 sources. It 
is clear
that there is a general line of deep infrared sources coming from the center of the
Mark 273 region in roughly the same direction (p.a. = ~235 deg.) as the 
Mark 273x/radio double and X-ray material (~220 deg.).  

\section{Radio sources around Mark 273}

Among the three closest radio sources to Mark 273 at 4.85 GHz (Becker et al. 
1991), there is a 42 mJy extended source 18 arcmin NE and a 27 mJy source 35
arcmin SW. A little further in this direction (46 arcmin) there is a radio source of
31 mJy which appears to coincide with the strong X-ray source of 49 cts/ks
mapped in Fig. 7. This same source is identified in Table 1 as a compact galaxy.
These outer radio sources lie closely along the NE-SW direction of the inner
radio sources Mark 273, 273x and dbl source SW (see Fig. 5).

Radio ejections from active galaxy nuclei are well accepted. X-ray ejections
from Mark 273 have been shown here. Physical association of X-ray
sources has been been shown in the past - for example six Seyferts, including 
Mark 273, were shown to have a factor 8 higher X-ray source density around 
them (Turner et al. 1993). A sample of 24 of the brightest Seyferts were shown to 
have paired X-ray sources across them, the great majority of which must be 
quasars or quasar-like (Radecke 1997; Arp 1997). The alignment of radio sources 
and X-ray quasars from these conspicuously ejecting ULIRG's would suggest,
therefore, that material in various different forms is ejected. 

We now see that for Mark 273 the infrared sources in Fig. 8 are distributed around 
a principal line of ejection. It raises the question of how they are related to the
ejection process and the other ejected material.

\section{Markarian 231}

     Fig. 9 shows that flanking Mark 231 there is a double radio source about 7.5
arcmin W and another double 2.2 arcmin to the E. (The latter has a faint 
companion to the NE making it actually a triple). Fig. 9b aligned directly below
Fig. 9, and to the same scale, shows that radio material is being ejected
toward the E from Mark 231 and that presumably the sources to the W represent
counter ejection in the other direction. The X-ray source to the east (left) is 
optically identified with a BSO and extended in the NE direction {\it as are also 
its radio contours}. 

     It is important to note that there is a faint but definite X-ray jet extending to 
the N from Mark 231. This is the same direction as the radio extensions to the N
and some small spots of radio emission as shown in Fig. 9b. Closely to the S of 
Mark 231
there is en extended radio source (Condon and Broderick 1998 ) and in the inner
60 milliarcsec there is a strong N-S triple source (Ulvestad et al. 1999; Taylor et
al. 1999). All of this is evidence for ejection and outflow of material in this
direction, a direction which only shifts by 5 deg. in p.a. from the innermost to the
outermost aligned sources (over 3 deg. distant for the infrared sources that we 
shall see later). 

      As we have seen in other cases here there is evidence for ejection 
in more than one direction from an active galaxy. In the case of Mark 231 the 
strong double radio source 7.6 arcmin slightly S of W is an obvious counter
ejection to the radio sources emerging from Mark 231 on the E.  The W source is 
conspicuous in that that it is a close double. (The sources on the E are also
closely multiple).

     The most striking object of all, however, is the {\it double X-ray source} 
which is aligned back through the double radio source to the nucleus of the
galaxy. An obvious interpretation is that as the X-ray source travelled outward
from Mark 231 it had its radio plasma stripped from it. This could occur in transit
of a homogeneous medium surrounding Mark 231, in conjunction with a discrete 
event of generation of radio plasma. Or it could be simply an encounter with a
discrete density enhancement in the medium.  

     It is apparent in Fig. 9 that at a slightly greater angle to the SW, there is 
another X-ray source with a radio source trailing at about the same distance 
behind it. In this and the previous case it is noticeable that the radio sources are
slightly above the line from the X-ray source back to Mark 231. This suggests
that the  medium which which stripped the plasma is rotating relatively slowly
counterclockwise. 

     Proof of such a mechanism would illuminate several puzzles about ejection of
matter from active galaxies. To this end we point to Fig. 9c which shows that the
double X-ray source is identified with a double quasar. (The brighter X-ray
source is catalogued as z = 1.272 and the optical image 52 arcsec E is
appreciably bluer implying a highly probable QSO of redshift to be determined).
The properties which support the relation of the double X-ray source to the
double radio source are then: 1) The relation of the stronger components of
the double to the weaker. 2) The similar orientation of the radio double, p.a. = 65
deg., and the X-ray double, p.a. = 75 deg. The separation of 52 arcsec of the
images in the X-ray and optical compared to the 24 arcsec separation of the
radio images would then imply a continuing separation of the X-ray images since
the event of the stripping of the radio plasma.

     Thanks to the detailed observations of Alan Stockton (astro-ph/9801056) we
can support this picture with observations of the radio galaxy 3C212. Fig. 10
shows that the optical objects f and g have passed out beyond the radio
material. The peculiar, identical shape of the radio contours and the optical
material leaves no doubt that the radio plasma has been removed suddenly from
the optical objects as they progressed away from the ejecting galaxy. It is also
revealing to note that the redshift of the material in the galaxy is z = 1.05 whereas 
the optical object, f, has z = .93. (Ejection velocities relative to the active objects,
and presumably their surrounding medium, have been shown to be of the order of 
.1c. (Arp 1998a)).       

     If this suggested mechanism of plasma stripping in fact operates, it furnishes
us with a general explanation of why the radio emission and the X-ray/optical 
objects can be generally but not exactly in the same positions. In fact we will see
later that in the famous radio galaxy, 3C31/NGC383, that strong radio filaments
trail toward the position of some high redshift galaxies and an even higher redshift
quasar. But the radio track is rotated somewhat from the X-ray/optical objects.     

     If the plasma generating events in quasars are intermittent it might help us
understand why, among similar appearing quasars, some are radio sources and
some are not. The higher percentage of quasars identified with X-ray sources 
than with radio sources could be explained by the fact that X-rays and optical 
synchrotron radiation would decay relatively 
quickly and therefore be much closer to the denser core of the quasar itself.
Periodic gas ejection events could also supply multiple clouds to account for
those quasars with multiple absorption line systems.

     Markarian 231 turns out to be particulary well suited to test these
suggestions because the lines of radio and X-ray ejection are so well marked.
Proceeding from the milliarcsec interior they appear fairly constant in direction
out to almost a degree on the sky. Perhaps this is due to the lack of rotational or
precessional angular momentum imparted by these particular ULIRG's, as 
suggested by their irregular morphologies.

\section{Surroundings of Mark 231 in the Infrared}

In the infrared field of Fig. 11 the long arrow shows the direction and distance of 
the 136 cts/ks X-ray source, 3C277.1, plus the spread of the radio ejections 
around this direction. The short arrow shows the extent of the radio and X-ray 
ejection N from Mark 231. The IR source is in the direction of the radio and X-ray 
ejection at p.a. =  243 deg. (The probable identification of this source is Y Uma
but there is a 1RXS, X-ray BSO within 23 arcmin.) There are also three sources 
aligned to the N at about p.a. = -10 deg., which is close to the inner radio and 
X-ray ejection to the N at about p.a. = -5 deg.

If these cold sources (they are perhaps even more prominent at 100 microns) are 
ejected from Mark 231 they confirm the result obtained in Fig. 8 for Mark 273. The
large angular separation of around 3 degrees on the sky, however would imply
that Mark 231 was much closer than the conventional redshift distance. Cold
infrared sources would also pose again the question of whether they were proto
quasars or by-products of the ejection.

\section{The X-ray field around Mark 231}

     Fig. 12 shows the distribution of the brightest X-ray sources around Mark 231
(14.3 cts/ks). This is the same elongation of radio sources and X-ray sources as
discussed in connection with Fig. 9. The strong, 136 cts/ks X-ray source which is
36 arcmin SW of Mark231 is optically identified with a relatively bright BSO at E =
 16.5 mag. as shown in Table 2. There is a moderately strong X-ray source, 
22.3 cts/ks at 50 arcmin to the NE of Mark 231 (not registered in Fig. 12) which is 
identified with a bright BSO of E = 17.4 mag. This is the most promising pair of
AGN's to be identified with the major ejection event of which we see the traces in
the interior. The 17.5 cts/ks X-ray source is the candidate double quasar shown
in Fig.9.

\centerline{TABLE 2}       
             
\centerline{Bright X-ray sources in Mark 231 field}

\vbox{\tabskip=0pt \offinterlineskip
\def\tr{\noalign{\hrule}}
\halign to \hsize{\tabskip=1em minus 1em&
\strut#\hfill&#\hfill&#\hfill&#\hfill&#\hfill&#\hfill&#\hfill
\tabskip=0pt\cr
 Cts &R.A.~~(2\rlap{000)}&\hfill Dec.&~~E&O-E&ID&Remarks\cr
\tr
  136 &12 52 26.2&~ 56 34 20 & 16.47 & 1.17 & BSO & 3C277.1\cr
  15.4 &12 54 49.1&~ 57 04 52 & 19.72 &  .47 & BSO &  \cr
  16.1 &12 54 51.5&~ 56 44 30 & 17.03 & 1.43 & Bcg & \cr
  17.5 &12 54 56.7&~ 56 49 42 & 19.21 & 1.53 & QSO & z = 1.272 & \cr
      "   &12 55 02.9&~~56 49 54 & 19.24  &   .65 & BSO & see Fig. 11 & \cr
   8.1  &12 55 24.7&~~56 56 14 &  19.53 &   .23  & BSO & \cr 
   4.9  &12 55 28.2&~~56 46 40 &  ~~ -  & ~ -   & ~- & no candidate & \cr 
  14.3 &12 56 14.2&~~56 52 25 &  13.84* & .84* & S1 & Mark 231 z = .041 & \cr
    4.2 &12 56 30.8&~~56 52 19 &  19.06 &  .44  & BSO & \cr  
    4.8 &12 56 48.3&~~57 03 45 &  19.77 &  .63  & BSO & \cr
  22.3 &13 00 33.5&~~57 28 35 &  17.35 &  .62   & BSO & \cr
  78    &13 00 43.3&~~56 21 28 &  19     & blue  & BSO & 30"S of brt star & \cr
 117   &13 00 52.1&~~56 41 05 &  16.27 &  .88  & BSO & possibly one or & \cr
     "    &13 00 54.5&~~56 41 11 &  19.03 &1.12  & BSO  & both are QSO's & \cr
}} 

* V and B-V

     Optical identifications will be discussed after Arp 220.

\section{Surroundings of Arp 220}

This ULIRG displays a chaotic morphology with strong dust absorption. Fig. 14  
shows that neutral hydrogen extends from the active galaxy down to
companions situated about 2 arcmin to the SW (Hibbard, Vacca \& Yun, 2000).

Fig. 15 shows that the brighter X-ray sources form a N-S string
in the shape of a shallow "S". These X-ray sources seem to be
ejected oppositely from the ULIRG with a small amount of rotation. The
material which formed the galaxies to the SW has apparently been ejected along 
this same path but  seems to have encountered resistance which has slowed its 
exit and allowed the development into fairly normal galaxies very close to the
ejecting ULIRG. In Fig. 14 the stream of hydrogen drawn out of the parent galaxy 
and ending exactly on the most active companion (an X-ray and radio source) is
evidence for the interaction which braked the normal escape of this material. 

      Redshifts of ejected compact material are, of course, initially high. We have 
seen ejected quasars and quasar-like objects in the preceding examples
of the active ULIRG's, Mark 273 and Mark 231. Similar candidates are apparent 
around Arp 220. Much previous evidence has indicated that these intrinsic 
redshifts decline as the objects evolve into more normal galaxies ( Arp 1998a,b). 
The only question is where they evolve, close to the galaxy due to interaction, or 
further out if the proto galaxy came out, for example, along the minor axis of an 
ordered galaxy.
 Empirically there has been evidence starting in 1970 that 
companion galaxies have systematically higher redshifts than their parent 
galaxies and this could only be reasonably accounted for by time dependent, 
intrinsic redshifts associated with their more recent creation.

    We will see another example, almost identical to the companions just
discussed for Arp 220, when we later consider briefly the active radio
source 3C31/NGC383. 

\section{Secondary ejection from the z = .09 galaxies}

     In order to measure the redshifts of the galaxies immediately SW of Arp 220
Ohyama et al. (1999) placed a one arcsec wide slit at an angle that passed
through the three largest companions. As Fig. 16 shows, this slit
serendipitously intercepted 4 much fainter galaxies on a line SE of the brightest 
companion. The redshifts of three of these objects could be measured from their
emission lines. They turned out to be z = .528, .529 and .523.

     Pursuant to conventional practice the investigators concluded that all
of these objects were accidental projections of objects at different distances. 
The picture was that of galaxies of z = .09 at 5x greater distance than Arp 220, 
with the higher z objects at a further 6x greater distance, all accidentally aligned 
closely in 
the same line of sight. However, if one computes the density of just the three faint 
objects in the 1 x 16.5 arcsec area of the sky in which they were found, one 
comes up with the astonishing figure of more than 2 million per sq. deg. Out to a 
radius of 37 arcsec from the galaxy B/C there should be over 800 such objects 
unless they are preferentially distributed along just the alignment of the 
spectrograph slit back to the large galaxy. It is instead attractive to consider that 
these z = .5 objects were ejected from the
active X-ray and radio galaxy at z = .09. They themselves are emission line
objects which makes the chance of their being accidental background galaxies
less (Arp 1982,p.62).  

The brighter galaxies at 
z =.09, in turn, are much too close to Arp 220 to be accidental. (At their apparent 
brightness of $m_R$ = $\leq$ 15.5 mag. about .03 galaxies would be expected 
this close to Arp 220 and, of course, very few would be X-ray sources). {\it
Moreover these 
otherwise morphologically normal galaxies would be of supposed quasar 
luminosities if placed at their redshift 
distances} (Ohyama et al. 1999 Table 2). Overall, considering the HI and X-ray 
connections, it would be difficult to avoid interpreting these z = .09 galaxies as 
having been ejected from Arp 220 and then having ejected even higher redshift 
objects. Empirically it seems to be another case of associated companion 
galaxies of various higher redshifts (Arp 1998a).

\section{The larger neighborhood of Arp 220}

Fig. 17 shows a larger field around Arp 220 in X-rays. It is clear that the
outermost X-ray sources, Nos. 1 and 10, continue the "S" shape from the
more interior regions pictured in Fig. 15. It is important to note that sources Nos.
2 and 9 (RSO and BSO in Fig. 15) are exactly aligned and almost exactly spaced
across the central ULIRG. The outer pair, Nos. 1 and 10 in Fig. 17, are also
exactly aligned and almost exactly spaced.

     The inner and outer pairs of X-ray sources therefore support the picture of
ejection with a slight counterclockwise rotation. The fainter sources interior to
the BSO source in Fig. 15 lead directly in along this spiral path, through the
companions at z = .09, and then into the ULIRG. The overall picture is then
one of material ejected from the active galaxy some of which develops into 
smaller objects of variously higher redshift. We next look at the outermost
regions surrounding Arp 220 in an  attempt to define over how large an area on 
the sky the association extends. To do this we process the full field of the 
archived PSPC observations as was
done earlier for Mark 273 and Mark 231. The results are shown in Fig. 18.

The most striking aspect of Fig. 18 is that the X-ray sources, particularly the faint 
ones, are distributed in long extensions to the NW and SE. The inner "S"
shape has turned over into an alignment across Arp 220 with a relatively small 
rotation of the inner sources, the same as observed in Figs. 15 and 17. One can
see the slightly smoothed background in Fig. 17 extending away on either side in
this large distribution of X-ray radiation.

Of considerable interest is the radio source 3C321. It is optically identified as a 
S2 galaxy of V = 16.0 mag.  Although one of the more intense radio 
galaxies in the sky it was not detected in X-rays by the Einstein IPC
(Fabbiano et al. 1984). In the ROSAT PSPC observation reduced here it is
clearly identifiable but too near the edge of the field (56 arcmin) to obtain a 
reliable flux measure. Note that 3C321 at z = .096 could represent material
ejected unimpeded at the same epoch as the z = .090 galaxies which were
stopped close to the SW edge of Arp 220.

As to the radio map of 3C321, two strong lobes are well aligned back toward Arp
220. Since there is no ready explanation for such an alignment one would
ordinarily assume it was accidental. We have seen, however, near alignments in
double radio sources near Mark 273, Mark 231 and along the X-ray ejection axis
from NGC2639 and NGC5985. (The latter two are not discussed here). 

That 3C321 
is physically associated with the apparent ejection alignment from Arp 220 is 
supported by the map in Fig. 19. There all the bright radio sources and all the
bright infrared point sources have been
plotted for a degree around Arp 220. The line through the ULIRG comprises 6 out
of the 8 radio sources in the area and 5 out of the 7 IRAS sources. Moreover that
line coincides closely with the line of X-ray sources, as can be seen by referring
to Fig. 18 directly above it.

\section{Bright X-ray sources in the Arp 220 field}

     It is very important to note the optical identifications of the X-ray sources
which make up this extended spiral alignment in the Arp220 field. Table 3 shows
that the strong source 20.3N is identified with a blue compact galaxy of relatively
bright apparent magnitude. The other strong source, 20.3S is identified with a
blue stellar object which is also unusually bright in apparent magnitude. These
two strong sources form a conspicuous pair across the active central galaxy. 

\centerline{TABLE 3}       
             
\centerline{Bright X-ray sources in the Arp 220 field}

\vbox{\tabskip=0pt \offinterlineskip
\def\tr{\noalign{\hrule}}
\halign to \hsize{\tabskip=1em minus 1em&
\strut#\hfill&#\hfill&#\hfill&#\hfill&#\hfill&#\hfill&#\hfill
\tabskip=0pt\cr
 Cts &R.A.~~(2\rlap{000)}&\hfill Dec.&~~E&O-E&ID&Remarks\cr
\tr
  20.3N &15 33 54.7&~ 23 56 15 & 16.34 &  ~~.93 & c.g.   & BSO's near & \cr
   ~5.0   &15 34 07.7&~ 23 29 39 & 18.28 & 1.19 & BSO & \cr
   ~8.9   &15 34 51.0&~ 23 46 05 & 20.65 &  ~~.73  & BSO &  possibly & \cr
  ~~ "    &15 34 51.8&~ 23 46 03 & 21.59 &  $<$1.59  & BSO & dbl QSO & \cr
   ~9.6   &15 34 55.0&~ 23 28 43 &~~   -   &  ~ -   & gal & SW plume & \cr
   ~7.3   &15 34 57.3&~ 23 20 12 & 13.88*&  ~~.88*  &  Sey & Arp 220 &\cr
   ~2.8   &15 35 02.3&~~23 15 29 &  15.93 &  ~2.98  & RSO & \cr 
   ~6.9   &15 35 06.1&~~23 36 56 &  19.61 &  ~2.18  & RSO  & \cr 
   ~4.2   &15 32 08.2&~~23 28 21 &  17.27 &  ~~.99  & BSO  & \cr
   ~9.4   &15 37 09.6&~~23 28 36 &  19.88 &  ~1.22   & BSO: & \cr  
  20.3S &15 37 14.5& ~~23 00 40 &  17.74 &  ~~.76   & BSO & \cr
}} 
* V and B-V

     As experience has shown, the outer members of ejected pairs tend to be
brighter, lower redshift, transitions between quasars and galaxies. (See
particularly Chu et al. 1998; Arp 1998b; 1999). It would therefore be predicted that
20.3N would spectroscopically turn out to be an active, moderately high redshift,
compact galaxy and 20.3S a moderately low red shift quasar. The inner quasar
candidates would be predicted to turn out to be fainter, higher redshift quasars. 

     In this respect it is interesting to note that in Table 3 the source at 6.9 cts/ks is
identified with a very red optical object. A ROSAT HRI position agrees with the
PSPC in identifying a very faint optical candidate. Together with its redness
it would suggest that the quasar candidate was highly dust obscured.  This in
turn would be highly interesting because it would suggest that the
famously heavy dust absorption in the central ULIRG extended to at
least 8 arcmin angular distance from Arp 220.

     Of course there are a number of further quasar candidates identified in Table
3. A line consisting of 9.4, 4.2 and 5.0 cts/ks sources goes roughly E-W. This is
consonant with ejection in more than one direction as observed in the cases of
Mark 273 and Mark 231 and other cases mentioned earlier. The pairs of quasar
candidates indentified in Tables 1 through 3 should be studied with the aim of
obtaining further empirical data concerning the speed, orientation and
evolutionary behavior of the ejected matter.

\section{Soft X-ray absorption from Arp 220}

When a column of hydrogen ($N_H$) containing X-ray absorbing metals 
intervenes, the soft
X-rays will be absorbed and the hardness ratio will approach HR1 = 1. As Fig. 20
shows, the sources around Arp 220 are conspicuously shifted toward this high
hardness ratio. The dashed line represents a control field around a bright star,
HR8905, which is at a comparable galactic latitude. The estimated excess in 
hardness ratio from Fig. 20 is about 0.4 in HR1. This translates into a difference in 
visual absorption of $A_V$ = .22 mag.

     In the detailed discussion in Appendix A we will argue that the actual
absorption is much greater than this minimum value.This means that in addition 
to absorption from our own galaxy at this
galactic latitude there is extra reddening and absorption from the material in the
environs of the ULIRG. {\it The surprising conclusion is that the
absorption extends out at least as far as 20 arc minutes and probably beyond 30
arcmin radius!}  Thus the gaseous component of the system extends over a 
degree in diameter on the sky. This angle is of the order subtended on the sky by 
larger Local Group Galaxies such as M33!

     We have seen individual, bright X-ray sources associated out to a diameter
approaching 2 degrees around Arp 220. But they may not be bound. The
absorbing gas, however, is much more likely to be travelling at less than the
escape velocity and this raises the question of what the system will evolve into - 
a large low redshift galaxy or a low redshift cluster of glaxies? In either case the 
empirical evidence on the apparent diameter requires the system to be much
closer than the redshift distance of the central, ejecting galaxy.

     The same situation applies to the ULIRG's Mark 273 and Mark 231. Although 
we do not show the HR1 off - axis plots here, it is readily apparent that they have
5 -9 sources which fall significantly above field sources at the same galactic
latitude and extend to radii of 20 arcmin and beyond. Such sources mark these  
systems as also extending to large angular diameters on the sky and thus closer 
than their redshift distances.

\section{Spectroscopic observations in Progress around Arp 220}

     A joint observing project on optically identified X-ray sources around active
galaxies is currently being carried out by E.M. Burbidge, Y. Chu and H. Arp.
Measures by Chu with the 2.2 meter Beiging telescope and E.M. Burbidge with
the 3 meter Lick Observatory reflector have yielded results some of which are in
the process of being reduced and reported. 

     In the case of Arp 220 it can be said that the two brightest X-ray sources in
the field are 20.3 cts/ks (20.3N and 20.3S in Table 3). The physical association of
this pair of X-ray sources is suggested by the fact that they are so much brighter
in X-rays than the remaining sources in the field and that they are almost exactly
equal in X-ray flux. These two have now been confirmed as quasars. they are of
low and rather similar redshift which would support the inference from their
diametric positions that they had been ejected in opposite directions from the
active nucleus.
     
     An even more exciting result is in the process of final reduction. This is the
pair of X-ray sources labeled RSO and BSO in Fig. 15. These two sources are
aligned as exactly as can be measured across the nucleus of Arp 220 and are
evenly spaced, each at about 7.3 arcmin distance. It is also apparent from Fig. 15
that there is a line of 4 or 5 X-ray sources which curves through the z = .09 
X-ray galaxies and lead directly to the southern member of the pair of sources,
designated BSO. 

     The candidate labeled RSO (red stellar object) is exceptionally red for a
quasar (O - E = 2.18 mag.) which makes it about 21.8 mag in the blue - very
difficult for a 3 meter telescope located on Mt. Hamilton above the lights of San
Jose. Nevertheless a 50 minute spectroscopic exposure recorded strong
emission lines which unmistakably signaled a quasar! Immediately thereafter the
southern member of the pair (BSO) was observed and it turned out to have less
strong lines but at very closely the same wavelength as the RSO. As a result we 
have a pair of quasars extremely well aligned across Arp 220 with closely the 
same redshift.

     Added to the pairing properties, of course, is the line of X-ray sources
connecting BSO back to Arp 220. The unusual redness of the northern
member of the pair may be related to the huge amounts of dust and absorption in
the ULIRG. If entrained material is drawn out in the process of ejection, the
quasar could be involved in a dense cloud of dust. The southern member is not so
red (O - E = 1.07) but this is still somewhat red for the usual quasar. X-ray
material, however,  seems to have been shed behind it in its outward track so 
more of the obscuring material may have been shed than for the RSO ejected in 
the other direction.

\section{Note in appendix which deals with absorption over the field associated
with ULIRG's and disturbed galaxies}
 
     Please see Appendix A for detailed discussion of the X-ray hardness ratios as
a measure of absorption and also some consequences of depressing the counts 
of more distant, background sources.

\section{Note added on quasar near Mrk 273}

     E.M. Burbidge has kindly allowed her spectroscopically determined redshift of
z = 1.168 for the 18.1 mag. BSO SE of Mrk 273 to be quoted. This is a particularly
important redshift because of its similar magnitude, strong X-ray flux and
close proximity to the previously known  z = .941 quasar. When corrected to the
reference frame of Mrk 273 ( z= .038) the new redshift goes to z = 1.089 and the
.941 redshift goes to z = .870. This means the redshifts fall very close to the
quantized redshift peak at z = .96 for quasars in general (Arp et al. 1990). In fact
one redshift differs by only +.066 and the other by -.055 from the z = .96 peak.
These are characteristic peculiar velocities relative to the parent galaxy (Narlikar
and Arp in preparation).

\section{Field of NGC6204}

     There is only a 5.2 ks PSPC exposure available for this fourth, bright ULIRG.
Consequently only the brightest X-ray sources can be examined (above 3.7
cts/ks). Fig. 22 shows that these X-ray sources are distributed principally along
directions NE and SW from NGC6240. 

     Inspecting the sources within 1 deg. radius around NGC6240 as compiled by
SIMBAD it is clear that also radio sources run approximately out to the NE
along this same line. In Fig. 21 we show in a gray scale map the disposition of
radio sources as recorded in the VLA continuum survey (nvss). Proceeding
outward from NGC6240 there is an extended radio source 5.4 arcmin to the east
of the ULIRG. It is low surface brightness and not readily identifiable with any
optical object. Probably it represents some radio plasma ejected in this direction.
Further out along a line to the NE is a string of four radio sources. Somewhat
above this line are four X-ray sources. The arrangement and spacing of the outer
three sources is similar - and reminiscent of the correlation between radio and
X-ray sources which we saw around Mark 231 (Fig. 9).

     Going in the other direction, we see three strong radio sources in a line to
the W. The striking feature here is that two of the sources are very close doubles.
We have seen a strong tendency in all of the previously mentioned active objects 
for the associated sources to be double. At such small separations they stand 
out from the rest of the radio sources in the field of Fig. 21 and are therefore
additionally indicated to be physically associated with NGC6240. The doubleness
might be attributed to their (recent) origin in a typical opposite ejection
from an ejected denser body which may or may not be in the near vicinity.

     The radio sources pictured in Fig. 21 are mostly confirmed in the 4850 MHz
surveys (Griffith et al. 1995) and the 365 MHz survey (Douglas et al. 1996). But
those fields extend further to the S than pictured in Fig. 21 and show a strong
radio source to
the SW which strengthens the appearance of a straight NE - SW line through
NGC6240 (as in the map only of X-rays shown in Fig. 22).The only three galaxies 
shown by SIMBAD in this field also fall
approximately along this line. The central of these three falls about 12 arcmin NE
of NGC6240, is 15.7 mag. and an IRAS source. Optical identifications should be 
made for these X-ray and radio sources and spectroscopic measures made also 
on the galaxies in order to study the details of their association.

\section{The bright radio galaxy 3C31/NGC383}

     Although NGC383 is not classified as an ULIRG (it only has a flux of 1.2 Jy at
100 microns), it is 1.5 mags brighter in the blue than Arp 220 and exhibits some of 
the same characteristics as the galaxies we 
have just been investigating.  In particular it is a strong X-ray source and has 
strong, ejected radio filaments which lead toward companion galaxies 
which are also X-ray sources. Curiously, it has an almost identical redshift to Arp 
220, z = .017 for NGC383 compared to z = .018 for the extreme ULIRG. 
NGC383 is, however, the 13th strongest radio E galaxy in the sky at 1420 MHz, 
and among that group, the 9th brightest apparent magnitude E galaxy (Arp
1968a). It has an emission lines which are ionized from the nucleus and probably 
shows a weak, broad emission component (Owen, O'Dea and Keel 1990).

     The most important feature for our purposes, however, is shown in Figs. 23
and 24 where it is seen that there is a tight group of galaxies which are strong in 
X-rays (1E 0104) and exhibit an X-ray tail pointing back toward NGC383 . The 
resemblance to Arp 220, as it was pictured in Figs. 13 -15, is remarkable . The 
startling fact then emerges that the redshift of the group associated with Arp
220 is z = .09, very similar to the redshift of the group associated with
NGC383 which is z = .11

     There is strong additional  evidence for ejection of X-ray objects in Fig. 23. 
Most of the discrete X-ray sources surrounding NGC 383 lie on
opposite ends of diameters passing close to the central galaxy. Most prominent 
is a line of sources at p.a. = 23 deg., including the 1E 0104 X-ray galaxy group.  
Both outer sources are elongated along this diameter. Fig. 23 also shows a pair 
of strong sources at p.a. = 100 deg. with material connecting back to the central
X-ray mass. A weaker pair is seen at p.a. = -45 deg. with the NW component a
double elongated back to the center. The innermost pair is particularly well 
aligned at p.a. = 60 deg. with the SW component exhibiting isophotes which 
conspicuously connect back to the central galaxy and the NE component being
a close double source. 

     The most significant aspect of these observations must be the X-ray 
filaments and isophotal extensions which lead back to the central source. Just
examining the X-ray map of the NGC383 surroundings, where the extensions fill a 
region out to 33 arc min radius, would seem to offer the
most direct demonstration possible of the ejection origin of these sources. (The
same kind of evidence is available from the galaxy cluster Abell 754 - see Arp
1998a, Fig. 7-17 and also Mark205 on the cover and Fig 1-7). The individual 
X-ray sources around NGC383 of course should
be optically identified as in Tables 1- 3 around the ULIRG's. Presumably they
would be mostly quasars whose properties could be usefully compared to
those around the ULIRG's.

\section{The connection between radio and X-rays}

     The radio map superposed on the X-ray map in Fig. 25 gives perhaps
the clearest evidence for the physical relation between the radio material
which is accepted as being ejected, and X-ray material for which similar evidence
has been accumulating. We notice that the radio filaments lead in the {\it
general direction} of the X-ray sources, but they deviate somewhat, principally
toward the end of the track. This is essentially the same result as gleaned from
all the preceding ULIRG's dicussed in this paper. In NGC383 the radio track is
more continuous and well marked but it is suggested that it is be due to one of 
the same two causes, or combination thereof:

     1) The X-ray sources and radio material are ejected along the same track but
the motion of the local medium separates the radio plasma from the denser
X-ray emitting bodies.

     2) Successive ejections are rotated and the radio track is an older remnant of
a preceding ejection. 

     In either case the radio and X-ray material should be superposable by simple 
rotations and translations. In NGC383 the southern extensions seem to be
well fitted with a small rotation. The northern radio extension seems to drift,
except for  a small spot, considerably  
westward at its end, perhaps due to some perturbation in the medium or local 
event. (For a purely translational drift note 3C212 in Fig. 10 of this paper). 

     A corollary to this mechanism is that if the radio and X-ray material try to
penetrate through appreciable material in the ejecting galaxy, the radio plasma
will be stripped without ever getting out. This would be a natural explanation of
why X-ray ejections seem to come out in several directions but radio ejections 
tend more to occupy one main channel. 

\section{Quasars associated with NGC383}

     There are some bright radio quasars and quasar like objects catalogued 
within a 
degree or two of NGC383 ( z = .603, 1.71 and a Sey1 with z = .015 which has  a 
companion of Z = .287). But the one for which it is possible to calculate a
very high probability of association is B in Fig. 24. It has z = 2.027 and falls 
{\it only 10 arcsec from A, the main galaxy in the group which has z = .11}.

     How probable is this to be an accident? In an objective prism survey Weedman
(1985) found quasars with  2 $\leq$ z $\leq$ 2.5 to have a density of .25/sq. deg. 
at a
magnitude of m$_{4500}$ = 19 mag. Quasar B has R = 18.9 mag. and therefore
would have a chance of approximately 6$x10^{-6}$ of falling accidentally within
10 arcsec of galaxy A. Moreover the analysis of Komossa and Boringher (1999)
make it seem likely that galaxy A and its group are the source of the strong 
X-rays. The association of an active X-ray galaxy with the quasar by chance is 
then vastly less
likely. It seems difficult to escape the conclusion that we have another example
of a physical association of a high redshift quasar with a low redshift galaxy
(Burbidge 1996; Arp 1998a;1999). It is to be noted that, although the quasar is
within the optical boundaries of galaxy A it shows no absorption lines at z = .11 
(Hewitt and Burbidge 1993). It shows a series of absorption systems from z = 1.97 
down to z = 1.75  but the absence of absorption lines from A would 
conventionally argue that it was not behind A.

    Of course we have argued that the origin of the group A galaxies was by
ejection from the central NGC383. This could make the quasar a secondary
ejection from an active, evolving companion or a later epoch proto galaxy
either entrained or traveling the same track. What then is the probability that A
has come from NGC 383, or for that matter that the quasar has come directly
from NGC383?

     Fig. 26 shows the X-ray contours obtained from the Einstein IPC. These
observations are with a less sensitive detector than used for the  ROSAT maps 
in Figs. 23 and 25 but at a somewhat higher energy, namely the .3 to 3.5 keV band.
It is clear that the isophotes of 3C31 are extended toward 1E 0104 and the
contours of the X-ray source are extended back toward 3C31. Allowing a
generous $\pm$ 10 degrees in pointing coincidence, the chances of accidental
mutual alignment are about 1 in 100. But now we have to ask what are the
chances a 3C radio source would be encountered in an area of 16.4 arcmin
radius. About 400 3C sources in about 30,000 sq deg. of sky gives another factor
of about 1 in 100. So the total chance of finding a galaxy as active as NGC383
with mutual X-ray alignments to 1E 0104 is about 1 in 10,000.

     But perhaps the most compelling evidence for association is the similarity
with Arp 220, where a tight group of X-ray galaxies of about the same redshift are
linked directly back to the ULIRG by X-ray and HI material. 

\section{Distances and quantizations of redshifts}

The angular distance from Arp 220 to its associated (z = .09) galaxy group is 2 
arcmin whereas it is 16 arcmin in the (z = .11) NGC383 case. The brightness of the 
NGC383 system makes it tempting to acount for this by arguing that NGC383 is 
nearer the observer. But we should remember that the 3C321 sytem at z = .096
was also associated with Arp 220 but at 56
arcmin radial distance. We argued there that the retarding interaction with the 
ejecting galaxy might determine the different distances traveled by the ejecta at 
their observed evolutionary stage.  

     The distances of these parent galaxies is difficult to estimate because their 
redshifts
appear to be an intrinsic property related to their age rather than a measure of
their distance. There are several ways of seeing this. One is through the evident
quantization of their redshifts. For example the redshifts of Mark 273 and Mark
231 are z = .038 and z = .041 respectively. The redshifts of Arp 220 and NGC383
are z = .018 and z = .017 respectively. The latter redshifts are very close to the
5000 km/sec peak in redshifts over the whole sky but particularly of the 
Perseus - Pisces filament that stretches almost 90 deg. across the sky. (see Arp
1987 p. 129ff; Arp 1998a p. 149ff). There is an obvious problem with interpreting
this as a shell of galaxies expanding away from our own in every direction. As
an alternative the
above references indicate that, empirically, galaxies at these redshifts give
evidence of being associated with brighter, more nearby galaxies.  In the case of
3C31/NGC383 it falls only 3.2 deg. away from the Local Group Seyfert 3, NGC404.

     It is also evident that quasars in these systems which have
measured redshifts come close to the global redshift peaks of z = .06, .30, .60, .96,
1.41, 1.96, 2.64 ...etc. For example the quasar B, SSW of NGC383 has z = 2.027. (In
the reference frame of NGC383 z$_Q$ = 1.976).
The intrinsic redshifts seem to decline in discrete steps as they evolve
toward more normal galaxies. It would be natural to reason that the ULIRG's and
other active galaxies are part of this evolutionary sequence and still have a
dominant component of the age related redshift which is quantized. For this
reason it might be better to estimate the distance from the apparent magnitude of
the quasars. In the systems we have been examining, the quasars are quite bright
in apparent magnitude. The quasar B again, for example, is  R = 18.9 mag. This is
bright for a quasar of this redshift and is similar to the properties of the line of
quasars coming SW from M33, another Local Group galaxy near NGC404 and
NGC383. (See Arp 1987 p.71ff).

     An other example of an object like those we have been studying is shown in
Fig. 27. It is the disturbed spiral IC1767. Amazingly it has a redshift of z = .0175,
just between that the .018 of Arp 220 and the .017 of NGC383. It has been argued
to be a southern extension of the Perseus-Pisces Cluster (Maurogordato, Proust
and Balkowski 1991). Like NGC383 it is only a modest infrared, IRAS source
of flux 1.0 Jy at 100 microns (but rising steeply toward longer wavelengths).
 
Like Arp 220 and NGC383, IC1767 shows evidence for ejecting radio sources and
higher redshift objects.
Its most outstanding feature is a pair of very strong radio sources 
aligned across it at 1.3 deg separation on the sky. This pair was so strong it was
identified 31 years ago (Arp 1968b). Subsequently the radio sources were
measured to be quasars, both very close to z = .6. (If one refers the redshifts to
the central galaxy they become z = .588 and .640, which, after allowing for about 
cz = .02 toward and away ejection velocity, gives {\it both} quasars very close to
the major quantization peak of z = .60. The combination of these properties with
the strength of the radio sources and closeness of alignment gives negligible
chance of accidental association of these quasars.  

The argument now centers on the properties of these z $\sim$ .6 quasars. they
are unusually bright in apparent magnitude, V = 17.09 and 16.40 mag.,
exceptionally bright in radio wavelengths, and unusually widely spread across the
central galaxy. All these properties argue for an unusually closeby system. Thus
it becomes another argument for the active objects like Arp 220, NGC383 and
IC1767 to belong to the Perseus-Pisces redshift peak, but for all these galaxies
to be much closer than their conventional redshift distance.

\section{Summary: Empirical Properties of Active Galaxies}

     The operational definition of an "Ultra Luminous" galaxy is one which
deviates strongly above the Hubble relation between redshift and apparent
magnitude. Since the class of galaxies we are investigating here differs in
spectroscopic and morphological characteristics from the one that defines the
Hubble relation, the latter cannot be used to define the distances or luminosities 
of the ULIRG's. In fact the class of active and high surface brightness galaxies -
Seyferts, compacts and quasars - defines a conspicuously steeper slope than
the Hubble slope (Arp 1968a; 1998a, Fig. 2 - 19), ruling out expansion velocity 
redshifts unless the luminosities are progressively adjusted to compensate.

     We have investigated here a sample of active galaxies with the most extreme
deviation from the Hubble relation. We have found them to be strong X-ray
sources, show vigorous ejection of radio material and generally disturbed 
morphologies. They do not look like the relaxed, symmetric galaxy forms of the 
most luminous nearby galaxies - whose distances do not depend on  
redshift. Characteristically the central active galaxy has physically associated
companions which are even more active and of various degrees of higher
redshift. The activity and morphological disturbances seem to be associated with 
lower rather than higher luminosity and there are many new cases presented
here of apparently younger companions with higher redshifts than their galaxy of
origin.

     Mark 273 at z = .038 shows X-ray and optical connections to an active 
companion object at z = .458. Optical and radio connections lead toward a pair of
optical and X-ray bright quasars one of which has z = .941. Many strong X-ray 
pairs across Mark 273 have been optically identified as almost certain quasar 
candidates. A line of 100 micron infrared sources extends over 20 arcmin from 
Mark 273. Their nature and relationship to Mark 273 is a mystery to be 
investigated.

     Mark 231 at z = .041 shows well marked radio ejections in two, roughly 
orthogonal
directions. X-ray sources are associated with these radio patches in such a way
as to suggest the radio plasma is in varying degrees of being stripped from them.  
{\it The close double nature of many ejected radio sources is again seen and
becomes a general characteristic of ejecta from these active systems.} A string of 
infrared sources stretches over 3 degrees from Mark 231 in the direction of a major
interior radio ejection from this active ULIRG! Eleven quasar candidates are
optically identified as BSO or Bcg candidates, many in pairs.

     Arp 220 at z = .018 has a group of X-ray galaxies at z = .09 linked to it by
neutral hydrogen extensions. Moreover they are a  part of a chain of X-ray 
sources emanating from the central ULIRG in a shallow spiral shape which has
knots paired closely across the central object. The interpretation would seem to
require ejection with rotation.  {\it The 50 x 50 arcmin field shows the inner "S"
shape bending over into an extended spiral distribution of X-ray sources of the
order of 2 degrees in diameter!} Radio and X-ray sources are shown to extend
along this same line and for the same distance. Heavy absorption over the Arp
220 area shows up in the hardening of the X-ray sources over an area of the order
of a degree in diameter. 

     NGC6240 at z = .024 shows in a short exposure, X-ray sources extending on a 
line either side of the the ULIRG. Radio surveys show lines of radio sources 
extending out farther than 40 arcmin. Double sources are again featured.

     The bright radio galaxy NGC383 at z = .017 shows pairs of X-ray sources
connected diametrically across it. The strongest pair includes a group of X-ray
galaxies at z = .11 . There is a quasar of redshift z =
2.027 only 10 arcsec from the central galaxy in this group and they are both
apparently associated with the central radio galaxy (3C31). The radio extensions
from 3C31 are the strongest of any of the cases and appear to be only slightly
displaced from the main X-ray ejections.

     {\it Note the similarity of the X-ray galaxies of z = .11 associated with
NGC383 to the  X-ray galaxies of z = .09 connected to Arp 220. In one of the
best marked ejection lines from a Seyfert galaxy, NGC3516 (Chu et al. 1998), the
last X-ray galaxy in the line has z = .089.} The latter confirms both the ejection 
origin of these galaxies and the quantized nature of their redshifts. It is
suggested that only the distances from the ejecting central galaxy change in 
accordance with the amount of their interaction as they exit.       

     An important final point should be that the preceding paper analyzes a more
or less complete sample of the brightest ULIRG's. Fainter examples would be
expected to follow somewhat the same patterns. In recent sample of faint
ULIRG's (Stanford et al. 2000) there is a marked tendency for the central objects,
pictured in the K band, to have diametric companions or two or three 
condensations emerging in a jet like configuration.

\section{What are the distances of these active central glaxies?}      

     If we cannot use the conventional redshift distances, then  what are the real 
luminosities of the systems we have been investigating 
in this paper? I see two possibilities of estimating non-redshift distances:

     1) The systems all seem to be ejecting high redshift quasars of the kind that
have been physically associated with nearby galaxies. If we view the quasar
redshift as signaling an age related phase in its evolution toward a normal galaxy
then it may be that at this phase all have similar luminosities (for example if
ejected masses are similar). Then we could judge by the apparent magnitude of 
the associated quasars what the distance was. A preliminary assessment seems
to indicate the quasars around the active galaxies investigated here are
somewhat brighter than those associated with the sample of Seyerts in Arp 1997. 
Those Seyferts were judged primarily to be in the Local Supercluster so the
active galaxies investigated here would be indicated to be somewhat closer.

     2) The angular size of the associations of material we have seen around the
ULIRG's in this paper are of the order of a degree radius or more. This would
class them with Local Group galaxies. One might argue that the outer sources
are escaping into the general field from the central galaxy and only within a much 
smaller radius does the material fall back in and then evolve into a smaller, more 
relaxed galaxy or group of galaxies. (This may finally offer a legitimate use for
merger calculations).

      The intriguing question then arises: Will Mark 273 and Mark 231 evolve into
systems like Arp 220 and NGC383 and what will the latter two systems evolve
into? NGC383 is firmly a member of the Perseus-Pisces chain. As such it
consists of redshifts between 4000 and 6000 km/sec and exhibits extensions of 
several degrees on the sky. If the redshifts continue to lower and the luminosities
increase as all these objects age they may develop into what we consider more
"local" groups and clouds of galaxies.

     There are two kinds of observations which could shed light on this question:

     1) There was an infrared,  NICMOS observation (Scoville et al. 1998) of the 
central 
regions of Arp 220 which discovered 8 possible star clusters (unresolved). The
authors calculated that at the redshift distance of Arp 220 these clusters would
have an absolute K magnitude of M$_K$ = -13.5 mag. This turned out to be 
brighter by 1.5 magnitudes than, for example,  any in NGC5128, the giant E 
galaxy also know as 
Centaurus A.  Now if we were to move Arp 220 a factor of 10 closer, that would
lower the luminosity of these objects to M$_K$ = -8.5, about that of the brightest
stars in a galaxy. Have stars been resolved in Arp 220? A factor of 10 closer than
its redshift distance would place the ULIRG about half way between our 
Local Group and our Local Supercluster.

     2) Quasar redshifts, as mentioned earlier, are "quantized". Empirically the 
spread around the preferred values is less for quasars associated with
low redshift galaxies (Arp et al. 1990). This has been interpreted as meaning that
the $\pm$ .1c ejection velocities carry some quasars out into the field, but those
that are captured by the ejecting galaxy must slow down, lose this .1c ejection 
velocity and assume more closely their intrinsic, sharply quantized redshift
values. When such data is available for the systems discussed here it may be
possible to  identify the gravitational diameter of the system and hence judge 
its distance by this angular size criterion

     In general it could be remarked that there is a quite well established, empirical
sequence of evolution from high redshift, compact, low luminosity quasars to 
brighter medium redshift quasars to compact, active and disturbed galaxies and 
finally to more
relaxed, normal glaxies (Arp 1998b). The intrinsic redshift drops in steps along 
this sequence and it is implied that the ULIRG's considered here are in a phase of 
still evolving toward lower redshift - they are evolving from quasars, and like
quasars, they are much closer, and less luminous than their conventional redshift
distances. If this is true, one piece of terminology which might be salvaged from
previous work is the
name ULIRG - except that it now would have to mean "Under Luminous" galaxy.

\section{Acknowledgements}

     I would like to thank the many observers who allowed use of their results, 
some of them unpublished, which enabled a comprehensive view of the
properties of these bright, very active galaxies.

\section{Appendix A} 

The ratio of hard to soft X-rays is HR1 = (A - B)/(A + B), where A is 
the flux between .4 to 2.4. keV and B is between  .07 and .4 keV.
Testing the behavior of the hardness ratio on a five assumed
homogeneous star fields showed that the hardness ratio varies with off-axis 
distance, crowding in the center of the field and source strength. The control
field HR8905 was chosen as:

1) the closest match to Arp 220 in all these characterisitics. (N$_H$ = .43 for 
Arp 220 and .50 x $10^{21}cm^{-2}$ for the star field). 

2) Only sources greater than 1.3 cts/ks were used and 

3) appreciably extended sources were excluded. 

Correcting for a difference of .07 in HR1 due to the different $N_H$ in the two 
fields gives an estimated excess in hardness ratio from Fig. 20 of about 0.4 in 
HR1. This translates into a difference in N$_H = .4 x 10^{21}$ in the hydrogen 
column and, using the relation A$_V$ = .56N$_H$(10$^{21}cm^-2)$ developed
by Predehl and Schmitt (1995), a difference in visual absorption of A$_V$ = .22 
mag. is obtained. Doubling this for the absorption through the entire cluster
gives $E_{B-V}$ = .15 and A$_B$ = .6 mag.

     This must be a minimum absorption for the material associated with Arp 220.
Some reasons are: 

1) When the hardness ratio approaches 1 it becomes an insensitive measure of 
the absorption (see Pietsch, Trinchieri and Vogler 1998, Fig. 1). 

2) If there are any foreground sources in Fig. 20 they would
lower the excess hardening attributed to Arp 220. 

3) Some objects associated with the ULIRG may be obscured by dense clouds 
and be below the detection level of the exposure. 

In general, however, the points in Fig. 20 spread around the mean by roughly 
$\pm$ .4 as if we were seeing some on the near side and some on the far side of 
the group.

     An absorption value from completely different type of measurement, the 
Balmer decrement of the emission line galaxy B in Fig. 13 gives $A_V$ = .51
$\pm$.1mag. (Ohyama et al. 1999). Galaxy B, attached to Arp 220 by X-ray 
material is surely near the center of the group which implies the absorption 
A$_B$ through the whole group of 1.4 mag. Since Arp 220 with its 
enormous absorption and reddening (Becklin and Wynn Williams 1987 estimate 50
magnitudes of visual absorption ot the center) lies also in
Fig. 20 very near HR1 = 1.0, it implies that that HR1's near this value signal much
larger absorptions than the minimums we have estimated. The important result of
Fig. 20 then becomes the fact that these high absorptions extend out past 30
arcmin radius. 

     It should also be noted that recent measures at 200 microns (Alton et al. 1998)
show colder dust extending to larger radii than 100 micron dust in resolved
galaxies. If this is true of Arp 220 its diameter would grow even larger and raise
the question of what is the role of this dust in the evolution of this system.

     One point of general interest is that with a minimum of $A_B$ = .6 mag.
absorption, the normal background count of X-ray sources must be appreciably
depressed. But with the larger absorptions indicated in the further discussion, 
background sources could be almost completely supressed 
and therefore almost all the sources actually belong to Arp 220. This is a point
which has not been considered when analyzing presumably distant quasars
which show absorption lines from intervening absorption clouds.

     For example, in 32 total fields searched for quasars behind elliptical galaxies
and clusters of galaxies Knezek and Bregman (1998) found 13 quasars out to
distances of r $\leq$ 13 arcmin with X-ray counts $\geq$ 10 cts/sec. The density
of these quasars reached from 5 to 8 per sq. deg. between 8 and 2 arcmin from
the center. The expected background density of X-ray quasars using their
detection success rate was about 1.4 quasars per sq. deg. This concentration
of quasars toward the E galaxies and clusters is illustrated in Fig. 28.

The above result addresses
a long standing claim that the observation of lower redshift absorption lines in
high redshift quasar spectra proved that the quasars were at their large, emission
line redshift distances. Of course the only thing that the observation proved was
that some quasars were not in front of the clusters. They could be inside the
cluster or, if they did not show absorption lines they could be in front, or behind 
but seen between clear
patches. The only decisive observation would be to see if they were more
numerous in the direction of the galaxies than the surrounding background.
The Knezek and Bregman observations seem to have inadvertently established
that. The point that the Arp 220 observations now make, however, is that
because of the absorption of the galaxy or cluster (as evidenced by the
absorption lines in the quasar spectra) {\it the background count of quasar density 
should be to some extent lower in the direction of the galaxy or cluster.} Thus
the result quoted above for an excess number in the direction of the galaxies is 
actually a lower limit because an undepressed background was used for 
comparison. 

     In the direction of clusters of galaxies "...the significant overdensity of 
background bright quasars ... on a scale of 10 arcmin..." was noted by Wu and
Fang (1996). Later these authors reexamined the possibility that the excess
could be due to gravitational lensing and concluded "...the quasar galaxy
association remains an unsolved puzzle in today's astronomy..." (Zhu, Wu and
Fang 1997).  It should be emphasized that evidence for absorption in the
cluster can only strengthen this conclusion.

Figure Captions:

Fig. 1 X-ray contours on a POSSII, R image of Markarian 273 and 273x 
from Xia et al. 1998

Fig. 2 Deep R image from the Univ. of Hawaii 88-inch telescope by John
Hibbard. Arrow points to Mrk 273x.

Fig. 3 Smoothed and contoured HRI X-ray image of Mark 273 and
Mark273x (72 arcsec NE) by Thomas Boller.

Fig. 4 Deeper, lower resolution PSPC X-ray image smoothed and
contoured by Thomas Boller.

Fig. 5 VLA, 20cm continuum map of Mark 273 by Min Su Yun.

Fig. 6 VLA radio continuum map from survey (nvss). Strong X-ray
sources are marked with x's, including a quasar of z = .941 and a BSO of E =
18.1 mag.

Fig. 7 X-ray photons in a 50 arcmin radius PSPC
field. Brighter sources from the standard detection algorithm are marked in
counts per kilosecond. Mark 273 is at center with 13 cts/ks and Mark 273x has 8.
The two strong, candidate X-ray quasars in Fig. 6 are marked 21, referred to in
text as 21N for North, and 38. The data on the labeled sources is given in Table
1. Note elongation of source distribution NE to SW through Mark 273 and also 
pairs of sources SE to NW. The picture is 50 arcmin on a side (1 sky pixel =
$1/2$ arcsec)

Fig. 8 A plot of all infrared sources of
greater than .5 Jy at 100 microns within a one deg. radius of Mark 273.

Fig. 9 (a) High resolution radio map of 10.4 x 20.7
arcmin field around Mark 231. From 20cm, VLA, FIRST survey. Notice the pairs of
double radio sources flanking the central galaxy. (b) At the same scale as above, 
a PSPC X-ray exposure (sources are dark) overlayed with contours of radio 
sources from the low resolution VLA survey. (c) Optical map of double X-ray 
source consists of z = 1.272 quasar and O = 19.9 mag. BSO at p.a. = 75 deg.
(nvss).

Fig.10 Radio contours are here superposed on an HST image of the
radio galaxy 3C212 (Stockton 1998) Images marked f and g are optical objects

Fig. 11 At an infrared wavelength of 60 microns, a 6.1 x 6.1 deg. IRAS survey 
field is pictured

Fig. 12 X-ray photons within about 50 arcmain from a PSPC exposure.
Cts/ks are labeled for brighter sources. Mark 231 has 14.3 cts/ks. Data for
sources are given in Table 2

Fig. 13 Optical identification of the four largest galaxies around Arp 220
which has z = .018. Ohyama et al. 1999

Fig. 14 HI contours at the redshift of Arp 220 (z = .018) leading 
down to a group of galaxies at z = .09. By J. Hibbard

Fig. 15 Hard X-ray band, .5 to 2.4 keV, smoothed, showing curved string of
sources leading down from Arp 220

Fig. 16 The z = .09 companions SW of Arp 220 with 4 small high redshift
objects circled. Ohyama et al. 1999

Fig. 17 Standard detection X-ray sources in a 43 x 43 arcmin field. No. 3 is
Arp 220. No. 7 is a star

Fig. 18 X-ray photons (broad band .11 to 2.4
keV, slightly smoothed) within about a 50 arcmin 
radius with brighter sources labeled in cts/ks. Arp 220 is in the center,
unlabeled, but with a count of 7.3 cts/ks. The z = .09 companion galaxies are
an extended source of 9.6 cts/ks merged with Arp 220 to the SW at his scale
(also unlabeled). 
The 8.9 and 6.9 sources are Nos. 1 and 2 respectively in Fig. 17, the 2.8
source is No. 10 in Fig. 17 and the source just above the 2.8 (unlabeled in this 
Figure) is No. 9 in Fig.17.
The radio source 3C321 and its two lobes are sketched in to scale.

The enlarged view in Fig. 17 illustrates how the sources and background
radiation in the central regions turn over from an approximately vertical "S"
shape to a spiral distribution elongated in the NW to SE direction in this map

Fig. 19 Plot of all radio sources $>$ 70 mJy at 4.85 Ghz (triangles) and all 
 infrared sources $>$ .7 Jy at 100 microns (filled circles) within 1 degree of Arp 
220. From surveys by Becker et al 1991; Moshir et al 1989

Fig. 20 Hardness ratio, HR1, of X-ray sources as a function of angular 
distance from Arp 220. The dashed line indicates averages from a control field at
similar galactic latitude

Fig. 21 The dark spots show strong radio sources from the VLA low
resolution survey (nvss). Small x's show X-ray sources from Fig. 22. Two of the
radio sources on a line W of the ULIRG are just discernible as very close
doubles. Dashed outline represents an extended radio source. It is 42.5 arcmin
from NGC6240 to the westernmost double

Fig. 22 A PSPC X-ray map around NGC6240 (radius $\sim$ 1 deg.) 
showing standard detection
sources with liklihood greater than 10. From S. Komossa, private communication

Fig. 23 A PSPC X-ray map at .5 to 2.0 keV of 3C31/NGC383. Sources pairing
across the center are indicated by their position angles. Adapted fom Komossa
and Boehringer 1999

Fig. 24 The galaxy group 16.6 arc min SSW of NGC383 (the + marks the
centroid of the the X-ray source 1E 0104). Galaxies A,C,E,F have redshift z = .11. 
B is a quasar with z = 2.027. Red image by Gioia et al. 1986. Insert at higher
contrast shows the quasar embedded in the image of the dominant galaxy

Fig. 25 The same X-ray map as in Fig. 23 but now with the radio
jet contours. Overlay of a .6 GHz radio map from Strom et al. 1983 by 
Komossa and B\"ohringer

Fig. 26 The X-ray map of NGC383/3C31 from the Einstein Laboratory IPC. 
Two stars have been removed from the figure published by Gioia et al 1986

Fig. 27 {Very strong radio sources form a pair across the disturbed spiral
IC1767. Their redshifts are close to the z = .60 quasar peak and the galaxy's
redshift is at the 5000 km/sec galaxy peak

Fig. 28 Density of quasars found near E galaxies and the centers of galaxy
clusters by Knezek and Bregman (1998). The background is calculated from the
density of X-ray sources with their quasar detection rates but with no
background count suppression from absorption by the central object

\end{document}